\documentclass{kluwer}    

\newdisplay{guess}{Conjecture}
\begin{document}              
\newcommand{\Ms}{\mbox{$\,M_\odot \ $}}
\def \aa#1#2   {{\it Astr. Astrophys. \/} {\bf Vol.~#1}, {pp.~#2}}
\def \apj#1#2  {{\it Astrophys. J. \/} {\bf Vol.~#1}, {pp.~#2}}
\def \mnras#1#2{{\it MNRAS \/} {\bf Vol.~#1}, {pp.~#2}}
\def \pasp#1#2{{\it PASP \/} {\bf Vol.~#1}, {pp.~#2}}
                    
\begin{article}
\begin{opening}
\title{Massive Star Mergers: Induced Mixing and Nucleosynthesis}
\author{N.\surname{Ivanova}}
\author{Ph.\surname{Podsiadlowski}}
\institute{Astrophysics, Oxford University}
\runningauthor{Ivanova, Podsiadlowski}
\runningtitle{Mixing and Nucleosynthesis in Massive Stars Mergers}
\begin{abstract}
We study the nucleosynthesis and the induced mixing during the merging
of massive stars inside a common envelope.  The systems of interest
are close binaries, initially consisting of a massive red supergiant
and a main-sequence companion of a few solar masses.  We apply
parameterized results based on hydrodynamical simulations to model the
stream-core interaction and the response of the star in a standard
stellar-evolution code.  Preliminary results are presented
illustrating the possibility of unusual nucleosynthesis and
post-merging dredge-up which can cause composition anomalies in the
supergiant's envelope.
\end{abstract}
\keywords{Stars: evolution; stars: nucleosynthesis}
\end{opening}
 
\section{Introduction}

Common envelope (CE) evolution is one of the most interesting
evolutionary stages of an initially close binary.  It occurs when the
two components of a binary system orbit inside an extended common
envelope which is not in synchronous rotation with the embedded
binary (Iben \& Livio, 1993).  A binary encounters a CE phase by one
of three suggested channels: (1) a dynamical instability (Darwin
instability) or a secular tidal instability (\opencite{Hut};
\opencite{Lai93}), (2) in nova systems because of the expansion of the
nova shell engulfing the companion, and (3) as a consequence of
dynamical mass transfer.  Once a common envelope has formed, the
secondary continues orbiting around the core of the primary within
this common envelope.  As the secondary is affected by various drag
forces due to its interaction with the envelope, the orbit of the
binary slowly shrinks.  The transfer of angular momentum from the
orbital motion to the envelope causes the spin-up of the envelope.
The final result of this slow spiral-in depends on how much of the
released orbital energy has been deposited into the envelope compared
to the binding energy of the envelope.  The deposited frictional
energy may drive expansion of the envelope, causing its partial or
complete ejection.  If most of the envelope is ejected, the system
survives as a close binary consisting of the core of the primary and
the secondary.  This evolutionary path provides the favourite channel
for the formation of short-period binaries with compact components.

Here we are primarily interested in the situation where the deposited
energy is not sufficient to eject the common envelope. Then the
spiral-in continues until the secondary starts to fill its own Roche
lobe and begins to transfer mass to the core of the giant.
Eventually, the binary will merge resulting in the formation of a
rapidly rotating single star.  During this merger, material from the
secondary forms a stream which emanates from the secondary and falls
towards the primary core, encountering an ambient medium with
increasing pressure and density.  This hydrogen-rich material may
penetrate to a depth of about $10^{10}\,$cm where the
temperature of the ambient matter is as high as a few $10^8\,$K.  This
deep penetration may have two major consequences: (a) the initiation
of hydrogen burning through the hot-CNO cycle which may provide a
neutron flux sufficient for efficient s-processing; (b) the
injection/generation of high-entropy material near the primary core
which may lead to the dredge-up of helium, eventually changing the
surface composition of the merger product.  In this contribution we
present some preliminary results of the evolutionary calculations
which involve the modelling of the merging phase.

\section{Modelling the binary merger}

Modelling the merging process requires consideration of many different
aspects. The first is the response of the envelope of the primary due
to the presence of the secondary. This includes modelling the
gravitational effect of the primary, the luminosity generated by the
accretion onto the secondary and the frictional luminosity due to the
differential rotation of the common envelope (see Meyer and
Meyer-Hofmeister 1979; Hjellming and Taam 1991; Ivanova, Podsiadlowski
and Spruit, 2000).  A second important aspect is how to describe the
gas flow from the secondary, in particular, the entropy of the stream
material and the depth to which this flow can penetrate into the core
of the primary.  This also requires the self-consistent determination
of the mass-loss rate from the embedded secondary into the common
envelope, the density of the stream and the change of this entropy due
to the interaction with the ambient medium (Ivanova, Podsiadlowski \&
Spruit, 2001, IPS)

In our evolutionary simulations we follow a simplified scenario for
the spiral-in, taking into account all of these effects.  To model the
deposition of hydrogen-rich material in the core of the primary, we
use the prescription developed in IPS: this recipe provides an
estimate for the penetration depth based on a modified Bernoulli
integral and the entropy change of the stream material for a
parameterized stellar structure, where the entropy change 
in the post-shocked material comapre to the initial state 
is calculated as

$$K_{\rm S}  = {\frac {P_{\rm s}} {P} } 
\left ( {\frac {\rho_{\rm s}} {\rho}} \right )^{\gamma}
= 1 + k \cdot \eta_{\rho}^{\gamma-1} 
\left ({\frac {M_{\rm int}} {M_{\rm ext }}} \right ) ^2 \ .
$$

\noindent
Here $\eta_{\rho}$ is the ratio of the stream density to the
density of the ambient matter, $M_{\rm int}$ and $M_{\rm ext}$
are internal and external Mach numbers of the stream, $\gamma$
is the adiabatic index of the stream material.
For our calculations we used the entropy change coefficients
in the range $0\div0.4$.

Special considerations need to be taken in the calculations of nuclear
reactions within the stream-core impact region, where a convective
zone can develop. The standard assumption in calculating nuclear
reactions in evolutionary codes is that nuclear processes in
convective regions occur on timescales which are much longer than the
mixing timescale and that material is mixed efficiently in convective
zones so that no abundance gradient can develop there.
In the case where hydrogen-rich material is injected
directly into the hot zone, the characteristic timescale of the
hot-CNO cycle can be comparable to the timescale of the convective
mixing.  A method which can properly treat nucleosynthesis in this
environment has been developed by Cannon (1993) for Thorne--$\dot{\rm Z}$ytkow 
objects. For our merger simulations we use this method directly
in the evolutionary code for stages starting with the merging itself
up to the moment when the post-merging chemical composition is
uniformly mixed over convective zones in the star, well after the
merger has been completed.

As a starting point for the merger calculations we use a binary
consisting of a 18\Ms primary and a 2\Ms secondary.  The primary at
the start of the spiral-in has been evolved to the stage where it has
a core of about 6.8\Ms (and a He-exhausted core of about 4.1\,$M_{\odot}$).
The secondary is then assumed to spiral-in inside the envelope of the
primary, and the merger phase starts when the immersed secondary
begins to overfill its Roche lobe.  Since a primary application for
this scenario is modelling the progenitor of SN 1987A, we adopted a
chemical composition consistent with the composition of young stars
in the LMC, with $X=0.71$ and $Z=0.01$. We chose a mixing-length
parameter $\alpha = 2$ and included convective overshooting and
undershooting by 0.25 pressure scale heights.

\section{Results and discussion}

In our preliminary runs we have calculated two models, one with no
entropy change in the stream and one with a high entropy change.  The
main differences between the two calculations can be explained in
terms of how the hydrogen-rich material is injected into the hot,
He-burning zone.  In the case of no entropy change, the stream
penetrates into the convective He-burning zone directly. This leads to
an immediate very energetic response with very efficient
nucleosynthesis.  This causes a rapid expansion of the core of the
primary and results in the widening of the secondary orbit and a
temporary interruption of the mass transfer.

If there is large generation of entropy in the stream, $k=0.4$, 
the stream cannot penetrate as deep and does not reach the
convective He burning zone at the start of mass transfer.  Since there
is no dramatic expansion of the primary core, mass transfer continues
steadily at an approximately constant rate. In the impact zone, a
high-entropy region is generated.  With time, a convective zone just
below the impact zone starts to develop.  Once created, this zone
slowly erodes the core, mixing hydrogen down into hotter layers. The
growth of this convective zone proceeds exponentially, very slow at
the beginning, speeding up steadily up to the moment where it connects
to the He-burning convective zone.  At that point, hydrogen is quickly
mixed throughout the He-burning zone. Unlike the previous case, the
primary at this stage has accumulated much more hydrogen in this zone,
which therefore leads to much more dramatic nuclear
burning when the hydrogen is mixed into the helium-burning zone.

There are many uncertainties which can affect and significantly change
these results.  The models discussed above have been calculated with
the assumption that there is significant convective undershooting.
Nevertheless, calculations which have been done for a smaller entropy
change ($k=0.2$) but without undershooting have shown a
behavior similar to the behaviour in the second model; but the
model with $k=0.4$ and no undershooting did not show any
mixing of hydrogen into the helium-burning zone, though the entropy
generated in the core neighborhood during the merger can still induce
mixing at a later stage. Another possible uncertainty is the speed of
mixing in the convective zones, as it affects the rate with which the
hydrogen-rich material is moving into the hot zone, which affects the
nucleosynthesis.

Our preliminary conclusion is that, during the merger, hydrogen-rich
material from the secondary may be able to penetrate into the
He-burning zone, though in the case of high-entropy generation in the
stream, it is likely to happen only if there is significant
undershooting.

\end{article}
\end{document}